\documentclass[3p,,preprint,12pt]{elsarticle}
\makeatletter\if@twocolumn\PassOptionsToPackage{switch}{lineno}\else\fi\makeatother

      \makeatletter
\usepackage{wrapfig}
\newcounter{aubio}
\long\def\bioItem{%
\@ifnextchar[{\@bioItem}{\@@bioItem}}

\long\def\@bioItem[#1]#2#3{
 \stepcounter{aubio}
 \expandafter\gdef\csname authorImage\theaubio\endcsname{#1}
 \expandafter\gdef\csname authorName\theaubio\endcsname{#2}
 \expandafter\gdef\csname authorDetails\theaubio\endcsname{#3}
}

\long\def\@@bioItem#1#2{
 \stepcounter{aubio}
 \expandafter\gdef\csname authorName\theaubio\endcsname{#1}
 \expandafter\gdef\csname authorDetails\theaubio\endcsname{#2}
}

\newcommand{\checkheight}[1]{%
  \par \penalty-100\begingroup%
  \setbox8=\hbox{#1}%
  \setlength{\dimen@}{\ht8}%
  \dimen@ii\pagegoal \advance\dimen@ii-\pagetotal
  \ifdim \dimen@>\dimen@ii
    \break
  \fi\endgroup}

\def\printBio{%
  \@tempcnta=0
   \loop
     \advance \@tempcnta by 1
     \def\aubioCnt{\the\@tempcnta}
     \setlength{\intextsep}{0pt}%
     \setlength{\columnsep}{10pt}%
     \newbox\boxa%
     \setbox\boxa\vbox{\csname authorDetails\aubioCnt\endcsname}
     \expandafter\ifx\csname authorImage\aubioCnt\endcsname\relax%
      \else%
       \checkheight{\includegraphics[height=1.25in,width=1in,keepaspectratio]{\csname authorImage\aubioCnt\endcsname}}
        \begin{wrapfigure}{l}{25mm}
         \includegraphics[height=1.25in,width=1in,keepaspectratio]{\csname authorImage\aubioCnt\endcsname}
        \end{wrapfigure}\par
      \fi
     {\parindent0pt\textbf{\csname authorName\aubioCnt\endcsname}\csname authorDetails\aubioCnt\endcsname \par\bigskip%
     \expandafter\ifx\csname authorImage\aubioCnt\endcsname\relax\else%
      \ifdim\the\ht\boxa < 90pt\vskip\dimexpr(90pt -\the\ht\boxa-1pc)\fi%
     \fi}
      \ifnum\@tempcnta < \theaubio
   \repeat
   }

\makeatother

\usepackage{natbib}
\usepackage{tabulary,xcolor}
\usepackage{amsfonts,amsmath,amssymb}
\usepackage[T1]{fontenc}
\makeatletter
\let\save@ps@pprintTitle\ps@pprintTitle
\def\ps@pprintTitle{\save@ps@pprintTitle\gdef\@oddfoot{\footnotesize\itshape \null\hfill\today}}
\def\hlinewd#1{%
  \noalign{\ifnum0=`}\fi\hrule \@height #1%
  \futurelet\reserved@a\@xhline}
\def\tbltoprule{\hlinewd{.8pt}\\[-12pt]}
\def\tblbottomrule{\noalign{\vspace*{6pt}}\hline\noalign{\vspace*{2pt}}}
\def\tblmidrule{\noalign{\vspace*{6pt}}\hline\noalign{\vspace*{2pt}}}
\AtBeginDocument{\ifNAT@numbers
\biboptions{sort&compress}\fi}
\makeatother

\usepackage{ifluatex}
\ifluatex
\usepackage{fontspec}
\defaultfontfeatures{Ligatures=TeX}
\usepackage[]{unicode-math}
\unimathsetup{math-style=TeX}
\else 
\usepackage[utf8]{inputenc}
\fi 
\ifluatex\else\usepackage{stmaryrd}\fi

\usepackage{url,multirow,morefloats,floatflt,cancel,tfrupee}
\makeatletter

\AtBeginDocument{\@ifpackageloaded{textcomp}{}{\usepackage{textcomp}}}
\makeatother
\usepackage{colortbl}
\usepackage{xcolor}
\usepackage{pifont}
\usepackage[nointegrals]{wasysym}
\makeatletter

\def\mcWidth#1{\csname TY@F#1\endcsname+\tabcolsep}

\def\cAlignHack{\rightskip\@flushglue\leftskip\@flushglue\parindent\z@\parfillskip\z@skip}
\def\rAlignHack{\rightskip\z@skip\leftskip\@flushglue \parindent\z@\parfillskip\z@skip}

\@ifundefined{etal}{}{}

\usepackage{ifxetex}
\ifxetex\else\if@twocolumn\@ifpackageloaded{stfloats}{}{\usepackage{dblfloatfix}}\fi\fi

\AtBeginDocument{
\expandafter\ifx\csname eqalign\endcsname\relax
\def\eqalign#1{\null\vcenter{\def\\{\cr}\openup\jot\m@th
  \ialign{\strut$\displaystyle{##}$\hfil&$\displaystyle{{}##}$\hfil
      \crcr#1\crcr}}\,}
\fi
}

\AtBeginDocument{%
  \@ifpackageloaded{endfloat}%
   {\renewcommand\efloat@iwrite[1]{\immediate\expandafter\protected@write\csname efloat@post#1\endcsname{}}}{\newif\ifefloat@tables}%
}%

\def\BreakURLText#1{\@tfor\brk@tempa:=#1\do{\brk@tempa\hskip0pt}}
\let\lt=<
\let\gt=>
\def\processVert{\ifmmode|\else\textbar\fi}

\@ifundefined{subparagraph}{
\def\subparagraph{\@startsection{paragraph}{5}{2\parindent}{0ex plus 0.1ex minus 0.1ex}%
{0ex}{\normalfont\small\itshape}}%
}{}

\newcommand\role[1]{\unskip}
\newcommand\aucollab[1]{\unskip}
  
\@ifundefined{tsGraphicsScaleX}{\gdef\tsGraphicsScaleX{1}}{}
\@ifundefined{tsGraphicsScaleY}{\gdef\tsGraphicsScaleY{.9}}{}
\def\checkGraphicsWidth{\ifdim\Gin@nat@width>\linewidth
	\tsGraphicsScaleX\linewidth\else\Gin@nat@width\fi}

\def\checkGraphicsHeight{\ifdim\Gin@nat@height>.9\textheight
	\tsGraphicsScaleY\textheight\else\Gin@nat@height\fi}

\def\fixFloatSize#1{}
\let\ts@includegraphics\includegraphics

\def\inlinegraphic[#1]#2{{\edef\@tempa{#1}\edef\baseline@shift{\ifx\@tempa\@empty0\else#1\fi}\edef\tempZ{\the\numexpr(\numexpr(\baseline@shift*\f@size/100))}\protect\raisebox{\tempZ pt}{\ts@includegraphics{#2}}}}

\AtBeginDocument{\def\includegraphics{\@ifnextchar[{\ts@includegraphics}{\ts@includegraphics[width=\checkGraphicsWidth,height=\checkGraphicsHeight,keepaspectratio]}}}

\DeclareMathAlphabet{\mathpzc}{OT1}{pzc}{m}{it}

\def\URL#1#2{\@ifundefined{href}{#2}{\href{#1}{#2}}}

\def\UrlOrds{\do\*\do\-\do\~\do\'\do\"\do\-}%
\g@addto@macro{\UrlBreaks}{\UrlOrds}

\edef\fntEncoding{\f@encoding}

\makeatother

\def\style#1#2{#2}

\newif\ifmultipleabstract\multipleabstractfalse%
    \makeatletter
\def\ead{\@ifnextchar[{\@uad}{\@ead}}
\gdef\@ead#1{\bgroup
   \def\_{\string\underscorechar\space}
   \def\{{\string\lbracechar\space}
   \def\textdagger{\string\textdagger\space}
   \def\texttildeapprox{\string\texttildeapprox\space}
   \def~{\hashchar\space}
   \def\}{\string\rbracechar\space}
   \edef\tmp{\the\@eadauthor}
   \immediate\write\@auxout{\string\emailauthor
     {#1}{\expandafter\strip@prefix\meaning\tmp}}
  \egroup
}
\gdef\emailauthor#1#2{\stepcounter{ead}
      \g@addto@macro\@elseads{\raggedright
      \let\corref\@gobble
      \eadsep\texttt{#1} (#2)
      \def\eadsep{\unskip,\space}}
}

\makeatother
  
\usepackage{float}

\begin{document}

\begin{frontmatter}

    \title{
  \textbf{A Novel Patent Similarity Measurement Methodology: Semantic Distance and Technological Distance}    
}
    
\author[affc30f26b6787f42c8bcaa58b9c25632ec]{Yongmin Yoo}
\ead{yooyongmin91@gmail.com}
\author[aff7978416db64a4d3c91456a9ffa0aabea]{Cheonkam Jeong}
\ead{cheonkamjeong@arizona.edu}
\author[aff078a6c89fc564e9b9f8cc3bd09e89273]{Sanguk Gim}
\ead{worj2uk@gmail.com}
\author[aff6cca0bdaa3894c26b96f05653ec716fa]{Junwon Lee}
\ead{asdl0320@gmail.com}
\author[aa09066fcc7f8]{Zachary Schimke}
\ead{zackschimke@arizona.edu}
\author[abe732a782131]{Deaho Seo\corref{contrib-fdacfacabdee4c0ea555f879faeba1ec}}
\ead{seo\_daeho@naver.com}\cortext[contrib-fdacfacabdee4c0ea555f879faeba1ec]{Corresponding author.}
    
\address[affc30f26b6787f42c8bcaa58b9c25632ec]{Department of Computer Science\unskip, 
    University of Auckland\unskip, Auckland\unskip, New Zealand}
  	
\address[aff7978416db64a4d3c91456a9ffa0aabea]{Department of Linguistics\unskip, 
    University of Arizona\unskip, Arizona\unskip, Tucson\unskip, United States}
  	
\address[aff078a6c89fc564e9b9f8cc3bd09e89273]{
    SR Universe\unskip, Seoul\unskip, Republic of Korea}
  	
\address[aff6cca0bdaa3894c26b96f05653ec716fa]{
    Royal Melbourne Institute of Technology\unskip, Melbourne\unskip, Australia}
  	
\address[aa09066fcc7f8]{James E. Rogers College of Law\unskip, 
    University of Arizona\unskip, Tucson\unskip, Arizona\unskip, United States}
  	
\address[abe732a782131]{School of Information\unskip, 
    Yonsei University\unskip, Seoul\unskip, Republic of Korea}

\begin{abstract}
Patent similarity analysis plays a crucial role in evaluating the risk of patent infringement. Nonetheless, this analysis is predominantly conducted manually by legal experts, often resulting in a time-consuming process. Recent advances in natural language processing technology offer a promising avenue for automating this process. However, methods for measuring similarity between patents still rely on experts manually classifying patents. Due to the recent development of artificial intelligence technology, a lot of research is being conducted focusing on the semantic similarity of patents using natural language processing technology. However, it is difficult to accurately analyze patent data, which are legal documents representing complex technologies, using existing natural language processing technologies. To address these limitations, we propose a hybrid methodology that takes into account bibliographic similarity, measures the similarity between patents by considering the semantic similarity of patents, the technical similarity between patents, and the bibliographic information of patents. Using natural language processing techniques, we measure semantic similarity based on patent text and calculate technical similarity through the degree of coexistence of International patent classification (IPC) codes. The similarity of bibliographic information of a patent is calculated using the special characteristics of the patent: citation information, inventor information, and assignee information. We propose a model that assigns reasonable weights to each similarity method considered. With the help of experts, we performed manual similarity evaluations on 420 pairs and evaluated the performance of our model based on this data. We have empirically shown that our method outperforms recent natural language processing techniques.
\end{abstract}
    
  \end{frontmatter}
    
\section{Introduction}
Patents, a major form of intellectual property rights, give inventors the legal right to protect their inventions for a limited period of time against infringement by others making, using, or selling the invention without permission.\cite{an2021improved} Due to the rapid development of technology, the number of patents files each year has increased significantly. Additionally, due to the development of convergence studies, the format of the technology itself is often a combination of various technologies rather than a single technology, making the technology itself very complex and very complex and difficult to understand. For this reason, patent analysis, which analyzes a large number of complex patents using various techniques and tools to discover insights from patents, is becoming increasingly important.\cite{abbas2014literature} Establish a patent road map and plan strategies through patent analysis\cite{yu2019obtaining,choi2013sao}, predict emerging technologies to predict future technologies\cite{joung2017monitoring,lee2018early}, or determine how much opportunity a technology has through technological opportunity detection. Analyzing patents, such as detection\cite{yoon2015technology,yoon2012detecting} or patent valuation\cite{trappey2019patent}, has a lot of value.

Patent similarity measurement is one of the fundamental components of patent analysis. This is because it can detect the risk of patent infringement, prevent plagiarism, and evaluate whether the invention meets the criteria for novelty and innovation.\cite{wang2019measuring} A patent document is different from ordinary documents because it is a document made up of legal terms that describe a unique technology. Additionally, a patent is a special document consisting of special bibliographic information such as title, abstract, inventor, references, international patent classification, and assignee. Therefore, analyzing patent documents and measuring the similarity between patents is a very challenging task. 

Research measuring the similarity of past patents has been conducted steadily for a long time. Numerous researchers have proposed methods to measure patent similarity in various aspects. A study proposed a method to measure the similarity between patents considering the technical aspects of the patent.\cite{an2021improved,leydesdorff2014interactive,hain2022text,park2017application,yan2017measuring,eisinger2013automated} Some studies also proposed to measure the similarity of patents by considering the bibliographic information of the patent\cite{choi2022two,chen2017patent,rodriguez2015new,rodriguez2016patent,sharma2017patent}, and a study that proposed a method to measure similarity between patents considering the semantic aspects of patents\cite{vowinckel2023searchformer,siddharth2022enhancing,ni2021similarity,liu2021patent}, there are many methods to measure the similarity of patents. However, each method has its own limitations. Studies that measure patent similarity by considering the technical aspects of patents mainly consider the IPC code of the patent. The IPC code for patents is a hierarchical classification system of language-independent symbols for patent classification, covering all areas of technology. It is the hierarchical structure that makes IPC codes suitable for measuring patent similarity. However, IPC is an ambiguous classification system in that it relies heavily on existing technologies and their combinations when defining new and emerging technologies, which can lead to significant uncertainty.\cite{zhang2016hybrid} Therefore, if the similarity of patents is measured based only on unreliable IPC codes, it may lead to inaccurate results in measuring patent similarity. 

Studies that measure patent similarity by considering bibliographic information of patents mainly consider patent citation information. In addition, it is a good source that citation information can serve as a proxy to assess measurement patent similarity with bibliographic coupling network or co-citation network.\cite{rodriguez2015new,xu2018overlapping}Patent Citation Analysis is easy to use and very simple. Additionally, similarity measurement through citations sometimes shows powerful performance. However, there are two major limitations. 
(1)Due to citation hysteresis, new patents tend to be cited less than existing patents.\cite{an2021improved} 
(2)Some patent databases may provide inaccurate citation information.\cite{yoon2011identifying} Therefore, if patent similarity is measured considering only the bibliographic information of the patent, it may lead to inaccurate results in measuring patent similarity.

Studies that measure patent similarity by considering the semantic aspects of patents mainly consider the content aspects of patents. In the past, research\cite{arts2018text,park2013patent,choi2013sao} mainly measured patent similarity using text mining techniques such as keyword extraction. However, this has the limitation of being highly dependent on the inventor's keyword selection and language style. Recently, due to the tremendous advancement in natural language processing technology, studies \cite{schneider1995guidelines,lanjouw2004patent} have been proposed to measure patent similarity using deep learning models. Thanks to the rapid development and high accuracy of natural language processing using deep learning, patent datasets also show high accuracy. However, it has the disadvantage of not considering the characteristics of the patent. In particular, since patents often describe complex scientific technologies, a complete semantic analysis of the domain is not possible. For example, special symbols such as chemical formulas are not easy to analyze using a natural language processing model using deep learning.

To overcome this problem, unlike previous studies, this paper proposes an approach to measure the similarity between patents by considering not only deep learning methods but also characteristics of patents. The aspect that considers the technical aspect of a patent is to conduct research on the degree of technology overlap by considering the IPC code. The aspects that consider bibliographic information include not only citation information but also information on the assignee and inventor of the patent. Lastly, in terms of considering the semantic aspect, we consider the transformer model and LLM model, which are currently showing the best performance in natural language processing and select the model with the best performance. We then propose an optimized formula based on the model that assigns reasonable weights to each similarity method considered above. We used the Pearson coefficient and Spearman coefficient as evaluation metrics to compare and measure the performance of the proposed model with a widely used natural language processing model. Our model showed better performance than traditional deep learning methods such as CNN, LSTM, and Bi-LSTM, and the recently most popular large language model BERT and LLaMa2 models.

The rest of this paper is organized as follows. Section 2 briefly reviews related work, our methodology for measuring patent similarities is described in more detail in Section 3. Then, a case study is conducted to demonstrate the advantages of this methodology in Section 4. Then Result in Section 5 and discussion in Section 6. The last section concludes this contribution.

\section{Literature research}
In the literature research part, methods for measuring patent similarity were described from three perspectives. This section describes the research have done previously.

\subsection{Technological Similarity} Measuring technical similarity between patents is a method of measuring similarity based on the technical similarity between patents. Numerous studies have been conducted in the past to measure technical similarities between patents. Leydesdorff et al. studied US patents (USPTO) based on the International Patent Classification Code (IPC code). They proposed a method of measuring similarity based on IPC codes when analyzing data.\cite{leydesdorff2014interactive} Yuan Fu et al. Additionally, patent similarity analysis was conducted based on IPC. In particular, the similarity of groups competing in patent technology was compared and analyzed based on patent classification codes.\cite{lanjouw2004patent} Eisinger et al. also conducted an analysis using IPC. In this study, they recognized the limitations of searching for patents and proposed a search system using IPC codes to overcome them.\cite{eisinger2013automated} Yan et al,. proposed a technological distance measurement method for patent mapping. A method of finding the optimal method was proposed through comparative analysis using 12 similarity measurement methods based on technological similarity.\cite{yan2017measuring}

\subsection{Bibliography Similarity}Measuring patent similarity through bibliographic information is a method of measuring patent similarity using the characteristics of a special document called a patent. Unlike other documents, a patent is a document that contains special elements such as title, abstract, claims, citation information, inventor, assignee, and IPC code. In general, similarities between patents is measured using patent citation information. Choi et al,. proposed a system that recommends patents after measuring the similarity between two patents through analysis of citation relationships between patents.\cite{choi2022two} He proposed a method to measure the similarity between two patents by considering the forward citations of each patent. Rodriguez et al,. proposed a vector space model to analyze citation similarity between patents. A patent similarity measurement method was proposed based on the idea that citation relationships expressed as vectors are easy to analyze for patents with similar structures.\cite{rodriguez2015new} A study by Chen et al. proposed a method of measuring the similarity between patents by confirming that there is a technical connection between each patent in patent citations and analyzing the similarity between patents through this.\cite{chen2017patent} He started with the idea that patents may be technically identical if they cite the same patent or are cited by the same patents.

\subsection{Semantic Similarity}Measuring patent similarity through semantic similarity goes beyond simple analysis of the outside of the text and identifies patent similarity through understanding the meaning inside the text. Vowinckel et al,. proposed a method to analyze the similarity of characters through vectorization of patent characters. The SEARCHFORMER they developed is a transformer-based model that proposes a method of identifying and analyzing the similarity of newly invented patents by using existing patent documents rather than comparing similarity between patent documents.\cite{vowinckel2023searchformer} Siddharth et al. propose a semantic similarity analysis method for patents that focuses on natural language processing and embedding. This is used for patent classification and proposes a method of measuring patent similarity based on the similarity measurement standard used in the analysis \cite{siddharth2022enhancing} Liu et al,. proposed a patent similarity analysis method based on subject classification and semantic similarity by finding classification and semantic similarity using the title and abstract of the patent.\cite{liu2021patent} The study by Sinha et al. proposed a method to measure patent similarity based on the similarity of patent concepts rather than word matching analysis. In particular, a method was proposed to consider  the contextual meaning of the entire patent content rather than the fragmentary use of words, phrases, and frequencies.\cite{schneider1995guidelines}

\section{Methodology}
Our method considers three aspects to measure patent similarity which are the technical, bibliographic, and semantic aspects of the patent. We then propose an optimized patent similarity measurement method based on a model that assigns reasonable weights to each similarity method considered above.

\subsection{Technological Similarity}
\bgroup
\fixFloatSize{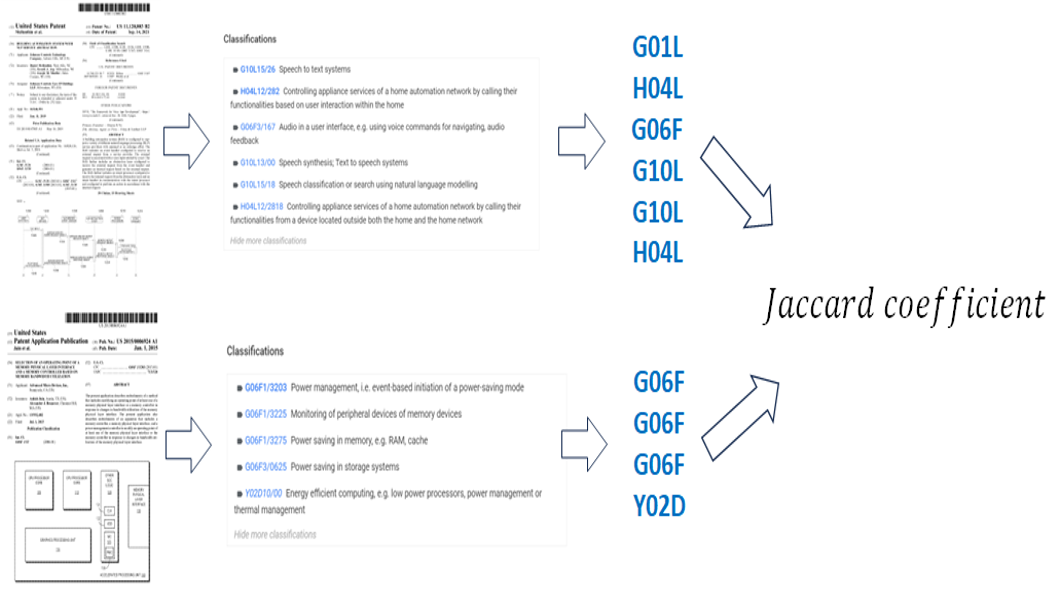}
\begin{figure*}[!htbp]
\centering \makeatletter\IfFileExists{imgs/technological.png}{\includegraphics{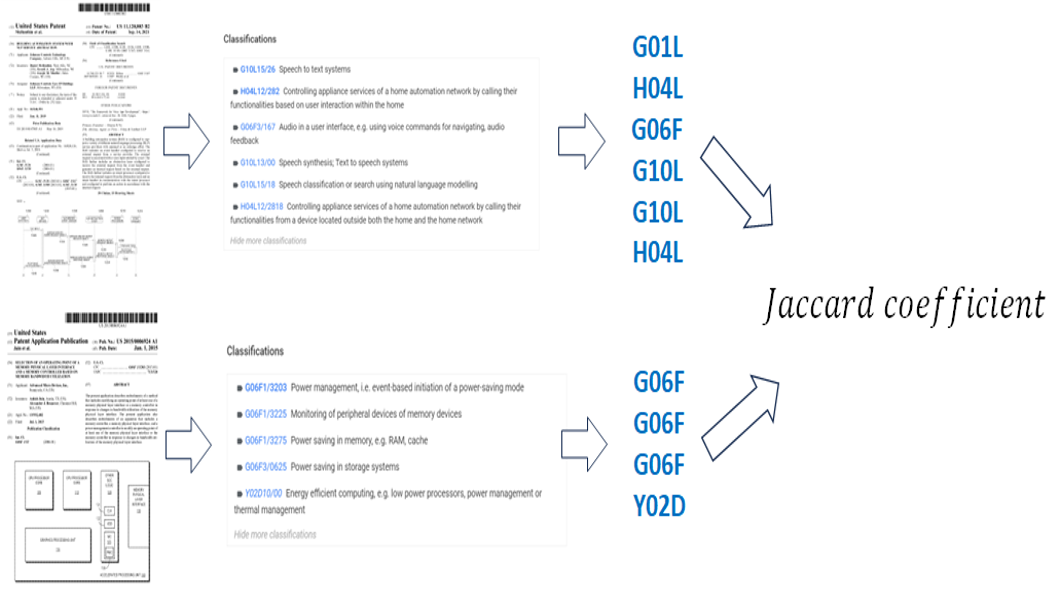}}{}
\makeatother 
\caption{{Model architecture for the technological similarity measurement}}
\label{f-b2059ef534e3}
\end{figure*}
\egroup
We first consider technical aspects to measure patent similarity. Similarity between patents is measured using international classification codes that classify patents by technology. The IPC code is a code that classifies detailed technologies and has a five-level hierarchical structure consisting of class, subclass, main group, and subgroup. Since it is very difficult to match all five levels, we check whether there is a match up to the main group, that is, the top three levels of the hierarchy. In other words, if the class, subclass, and main group are the same, it is considered a matching technology. Afterwards, we use the Jaccard coefficient of the IPC code to calculate the technical similarity between two patents with a score between 0 and 1. Technological Similarity model architecture is shown in Figure1 and Eq (1).

\let\saveeqnno\theequation
\let\savefrac\frac
\def\dispfrac{\displaystyle\savefrac}
\begin{eqnarray}
\let\frac\dispfrac
\gdef\theequation{1}
\let\theHequation\theequation
\label{disp-formula-group-f758c954fd9f472b8f6a701734878dc6}
\begin{array}{@{}l}\style{font-size:12px}\mathit{Technological\ Similarity} = \mathit{Jaccard}\left(\mathit{IPC}(\mathit{Patent}_A, \text{Patent}_B)\right)
\end{array}
\end{eqnarray}
\global\let\theequation\saveeqnno
\addtocounter{equation}{-1}\ignorespaces

\subsection{Bibliography Similarity}
\bgroup
\fixFloatSize{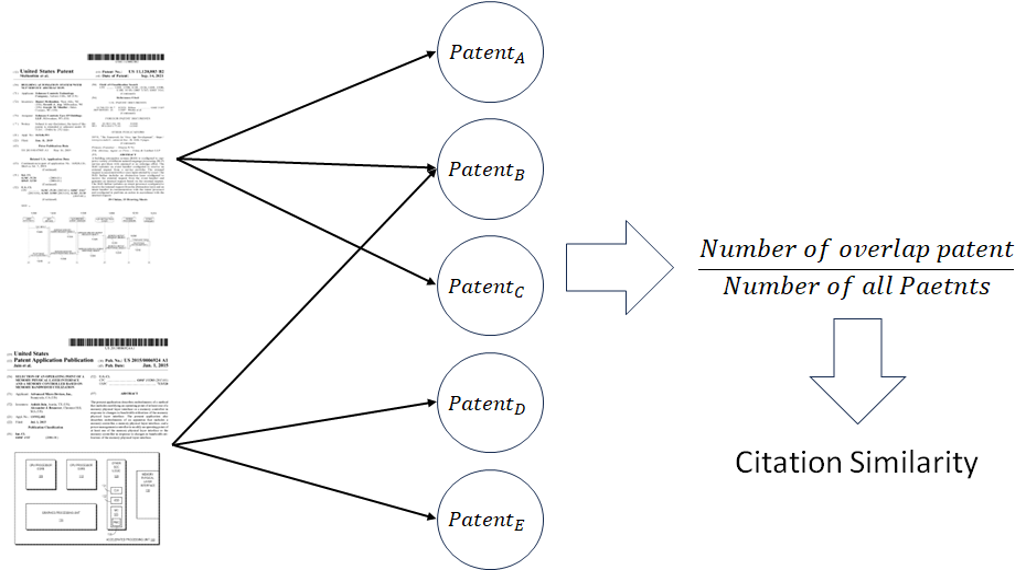}
\begin{figure*}[!htbp]
\centering \makeatletter\IfFileExists{imgs/Citation.png}{\includegraphics{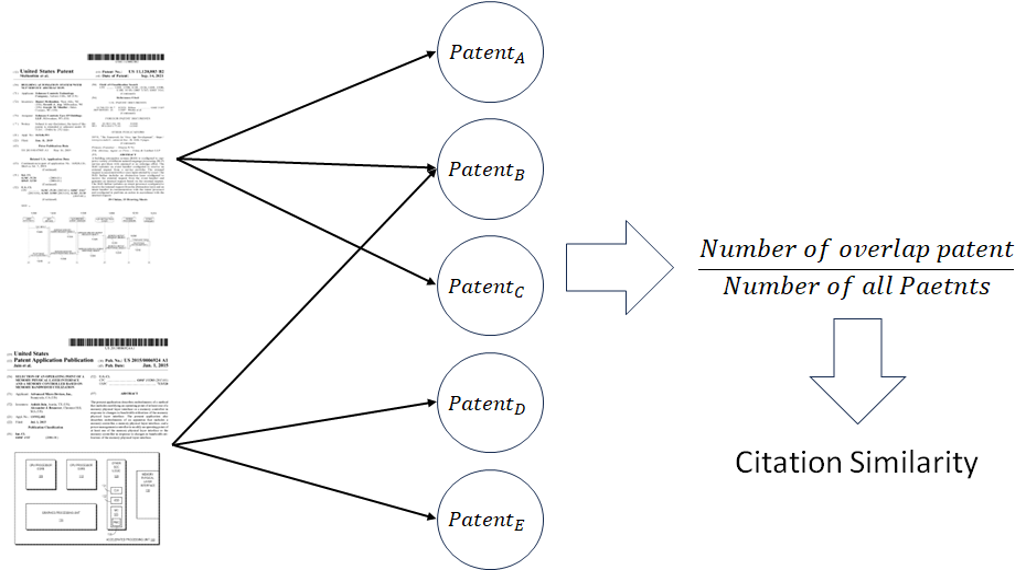}}{}
\makeatother 
\caption{{Model Architecture for the Citation Similarity}}
\label{f-608b9df0c687}
\end{figure*}
\egroup
We are considering a method, using the bibliographic information of patents to measure patent similarity. We measure the similarity between patents using citation information, inventor information, and assignee information, which are special elements of patents. First, we measure the similarity between patents using citation information between patents. Patent citations use backward citation information in which my patent cites other patents. In the case of a backward citation, it refers to a document submitted or cited by the examiner or applicant during the examination of the relevant patent application.\cite{lanjouw2004patent} We start from the idea that patents that cite the same patent may be technologically similar. Therefore, we measure the degree of citation of the two patents as shown in the Eq (2) below.

\let\saveeqnno\theequation
\let\savefrac\frac
\def\dispfrac{\displaystyle\savefrac}
\begin{eqnarray}
\let\frac\dispfrac
\gdef\theequation{2}
\let\theHequation\theequation
\begin{array}{@{}l}\style{font-size:12px}\mathit{Citation\ Similarity} = \mathit{Jaccard}\left(\mathit{Backward\ Citation}(\mathit{Patent}_A, \text{Patent}_B)\right)
\end{array}
\end{eqnarray}
\global\let\theequation\saveeqnno
\addtocounter{equation}{-1}\ignorespaces 

\bgroup
\fixFloatSize{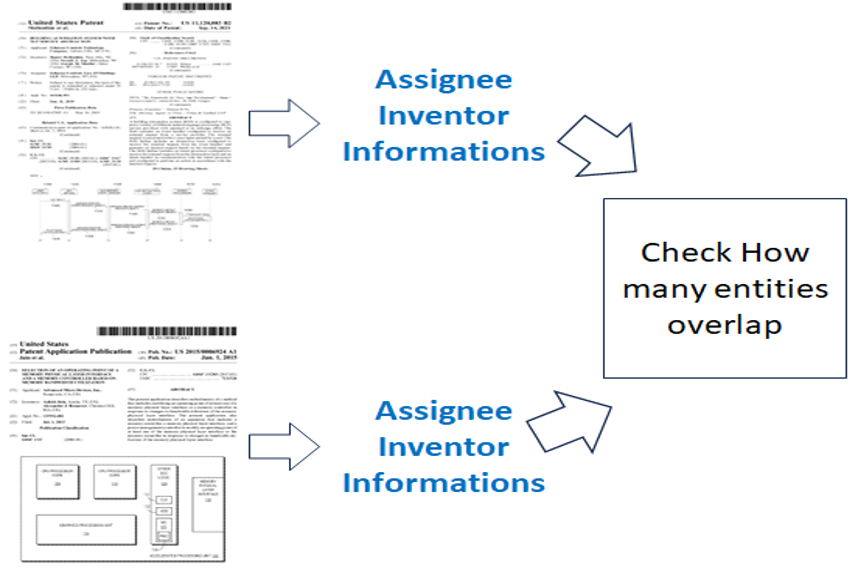}
\begin{figure*}[!htbp]
\centering \makeatletter\IfFileExists{imgs/Bibliography.png}{\includegraphics{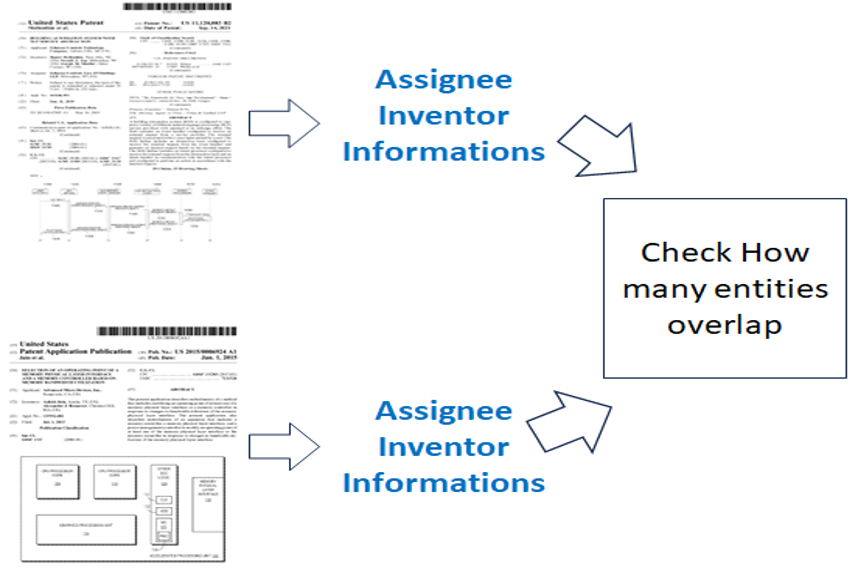}}{}
\makeatother 
\caption{{Model Architecture for the Bibliography Similarity}}
\label{f-608b9df0c687}
\end{figure*}
\egroup

Second, we propose a method to measure the similarity of patents using inventor information and assignee information. It starts with the idea that two patents will be technologically similar if they come from the same inventor. Additionally, we start with the idea that if the acquired company is the same, it may be technologically similar. In the case of inventors, the areas in which an individual can invent are very limited, so if a patent is applied for by the same inventor, 5percent weight is given to the number of matching inventors. In the case of an assignee, there may be a technical connection between the transferred patents, but since patents of various technologies are often considered, a weight of 3percent is given when the patent has the same assignee. The two equations are in Eq (3) and Eq (4) below.

\let\saveeqnno\theequation
\let\savefrac\frac
\def\dispfrac{\displaystyle\savefrac}
\begin{eqnarray}
\let\frac\dispfrac
\gdef\theequation{3}
\let\theHequation\theequation
\begin{array}{@{}l}\style{font-size:12px}\mathit{Inventor\ Similarity} = \mathit{number\ of\ Inventors} \times 0.05
\end{array}
\end{eqnarray}
\global\let\theequation\saveeqnno
\addtocounter{equation}{-1}\ignorespaces 

\let\savefrac\frac
\def\dispfrac{\displaystyle\savefrac}
\begin{eqnarray}
\let\frac\dispfrac
\gdef\theequation{4}
\let\theHequation\theequation
\begin{array}{@{}l}\style{font-size:12px}\mathit{Assignee\ Similarity} = \mathit{number\ of\ Assignee} \times 0.03
\end{array}
\end{eqnarray}
\global\let\theequation\saveeqnno
\addtocounter{equation}{-1}\ignorespaces

If the number of inventor value is 0, the value is entered as 1 in the equation below. Similarly, if the number of Assignee value is 0, it is entered as the value 1 in the equation below. The final equation for bibliographic similarity follows Eq (5) below.

\let\savefrac\frac
\def\dispfrac{\displaystyle\savefrac}
\begin{eqnarray}
\let\frac\dispfrac
\gdef\theequation{5}
\let\theHequation\theequation
\begin{array}{@{}l}\style{font-size:12px}\mathit{Bibliography\ Similarity} = \mathit{Jaccard}\left(\mathit{Backward Citation}(\mathit{Patent}_A, \mathit{Patent}_B)\right) \\
\quad \times (\mathit{number\ of\ inventors} \times 0.05) \times (\mathit{number\ of\ assignees} \times 0.03)
\end{array}
\end{eqnarray}
\global\let\theequation\saveeqnno
\addtocounter{equation}{-1}\ignorespaces

\subsection{Semantic Similarity}
\bgroup
\fixFloatSize{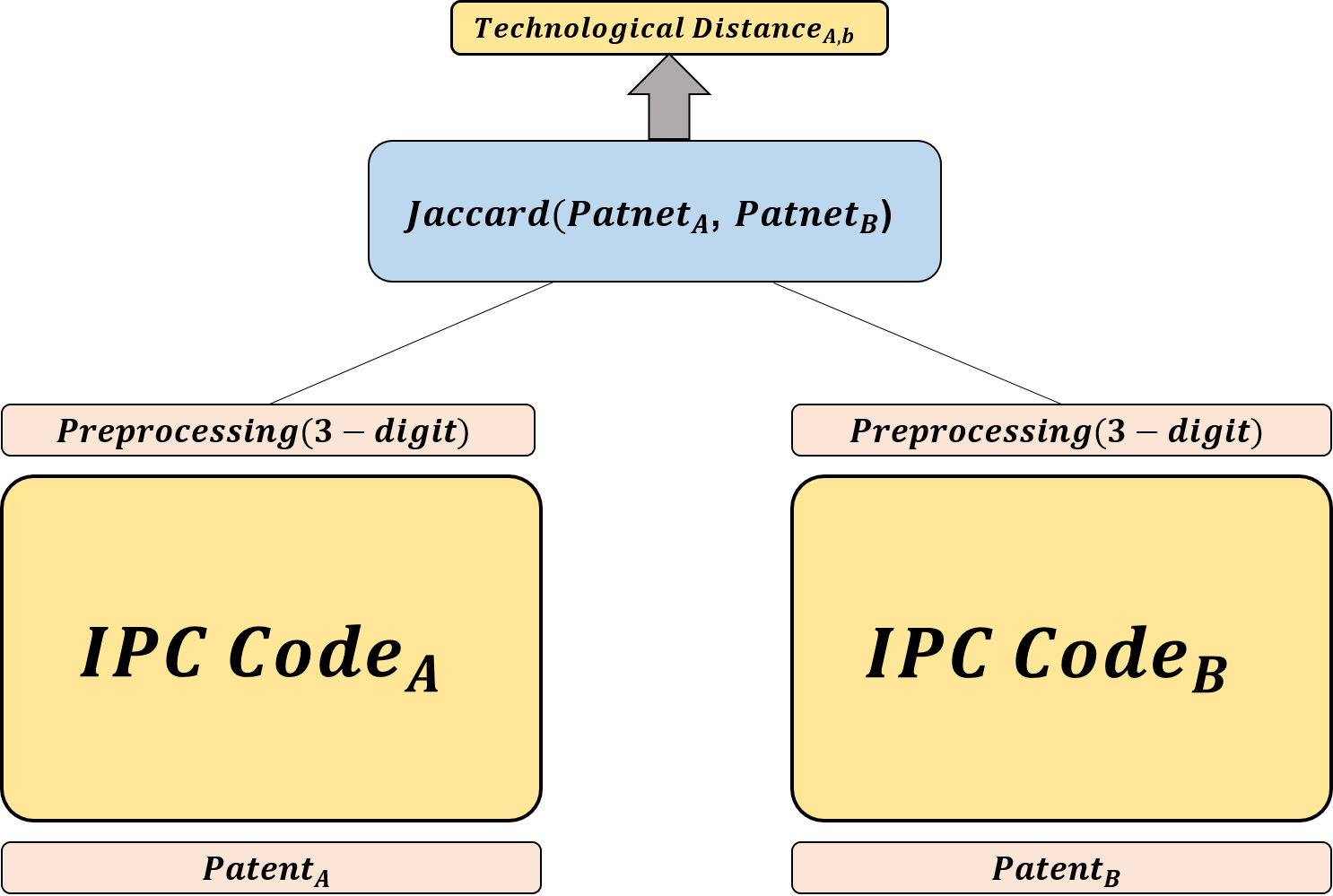}
\begin{figure*}[!htbp]
\centering \makeatletter\IfFileExists{imgs/model2.png}{\includegraphics{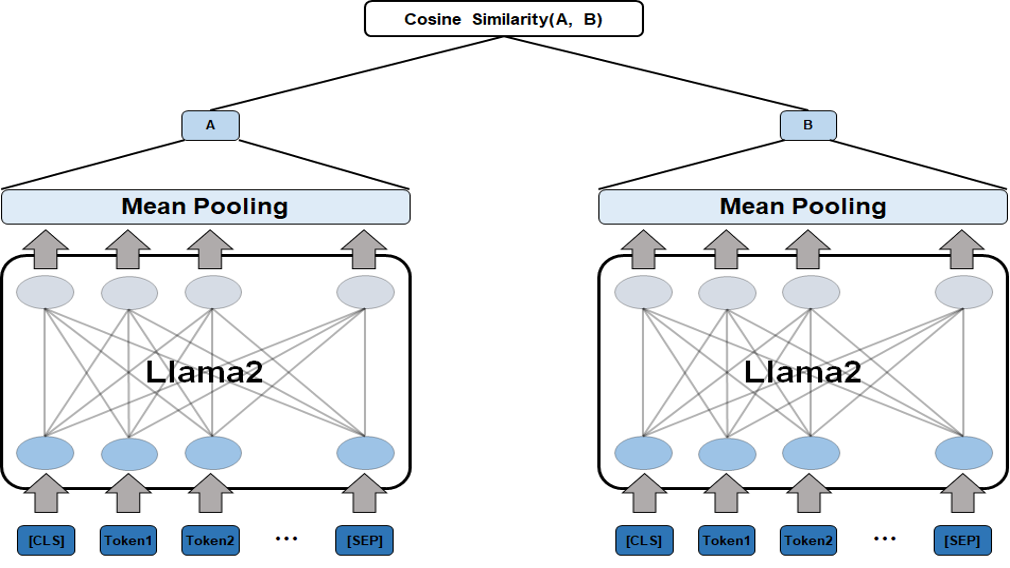}}{}
\makeatother 
\caption{{Model Architecture for the Semantic Similarity}}
\label{f-608b9df0c687}
\end{figure*}
\egroup
To measure patent similarity, we consider the semantic features of patents. We measure the similarity between patents using Llama2, which is the method with the best performance among natural language processing techniques.\cite{touvron2023llama} Llama2 is a Large Language Model(LLM) released by Meta in July 2023 and has more parameters than existing language models. Llama2 has sizes of 7B, 13B, 34B, and 70B, and in this study, Llama2 7B was used due to limitations in computing resources.

\subsection{Hybrid similarity}We will derive the most optimized formula by hybridizing the technical similarity of the previously proposed patent, similarity using bibliographic information, and semantic similarity. We propose an optimization method that weights the similarity through technical similarity and bibliographic information based on semantic similarity. We use actual data to find the optimal values of n and m that can produce the best performance. Our final equation follows the basic Eq (6) below.

\let\savefrac\frac
\def\dispfrac{\displaystyle\savefrac}
\begin{eqnarray}
\let\frac\dispfrac
\gdef\theequation{6}
\let\theHequation\theequation
\begin{array}{@{}l}\style{font-size:12px}\mathit{Proposed\ Method} = \mathit{Semantic\ Similarity}  \times \left(\frac{\mathit{Technological\ Similarity}}{n}+1\right) \\ \times \left(\frac{\mathit{Bibliography\ Similarity}}{m}+1\right)
\end{array}
\end{eqnarray}
\global\let\theequation\saveeqnno
\addtocounter{equation}{-1}\ignorespaces 

\section{Experiment}
In this section provides the dataset, evaluation method, and experimental results. In training the models, we use Python version 3.10.9 and Pytorch version 1.13.1 from Seoul, Korea. The computer specifications used in the experiment are as follows: 4 GeForce RTX 2080Ti and 16 Intel(R) Core(TM) i7-9800X CPU @ 3.80GHz.

\subsection{Dataset}Google patent is a platform where patents from 2016 to 2020 granted by United States Patent and Trademark Office are provided. Of them, Artificial Intelligence Patent Dataset \cite{giczy2022identifying} are adopted for this research. Using Google patent, we randomly extract 420 pairs patents from 2019 to 2020, with which we perform expert validation.

The expert panel consists of three scientists, each of whom has expertise in Data Analytics, Data Mining, and Artificial Intelligence, respectively. Before asking them to rate the patents, we set a guideline for rating, following advice by the author Zachery Schimke, a Juris Doctor candidate, who specializes in Patent at the University of Arizona. To be specific, we adopt a seven-scale rating scheme; it consists of the five odd numbers between 0 and 10, to which 0 and 10 are also added as exceptional cases. Then, we establish five criteria: Area, Task or Purpose, Methodology, and Application. Depending on the number of the overlapping criteria, rating scores are given, as described inTable~\ref{tw-80f871cf32bf}.

\begin{table*}[!htbp]
\caption{{Criteria for patent rating}}
\label{tw-80f871cf32bf}
\def\arraystretch{1}
\ignorespaces 
\centering 
\begin{tabulary}{\linewidth}{p{\dimexpr.16199999999999996\linewidth-2\tabcolsep}p{\dimexpr.838\linewidth-2\tabcolsep}}
\tbltoprule \cAlignHack \textbf{Score} & \textbf{Criterion}\\
\tblmidrule 
\rAlignHack 0 &
  Totally different\\
\rAlignHack 1 &
  1 similar property\\
\rAlignHack 3 &
  2 similar properties\\
\rAlignHack 5 &
  3 similar properties\\
\rAlignHack 7 &
  3 similar properties \& 1 somewhat similar property\\
\rAlignHack 9 &
  4 similar properties, but not the same\\
\rAlignHack 10 &
  Exactly same\\
\tblbottomrule 
\end{tabulary}\par 
\end{table*}
Following the guideline, each rater assesses how semantically similar two patents are solely based on patent abstracts with titles. In order to ensure reliability, we winnow out the data based on a variance analysis and then ask the author Zachery Schimkem, a law expert, to rate them. As described inTable~\ref{tw-ea4d982cc503}, we calculate \textit{Distance} and then set \textit{Threshold} as 8. If \textit{Distance} is larger than \textit{Threshold}, we separately store the data for \textit{Law expert rating}; otherwise, we calculate the mean score.

\begin{table*}[!htbp]
\caption{{Data rating flow} }
\label{tw-ea4d982cc503}
\def\arraystretch{1}
\ignorespaces 
\centering 
\begin{tabulary}{\linewidth}{L}
\tbltoprule 
$\style{font-size:14px}{\begin{array}{l}\boldsymbol L\boldsymbol a\boldsymbol w\boldsymbol\;\boldsymbol e\boldsymbol x\boldsymbol p\boldsymbol e\boldsymbol r\boldsymbol t\boldsymbol\;\boldsymbol r\boldsymbol a\boldsymbol t\boldsymbol i\boldsymbol n\boldsymbol g\boldsymbol:\\Asking\;a\;law\;expert\;to\;rate\;patents\\if\;scores\;rated\;by\;the\;expert\;panel\\have\;huge\;discrepancies\\\\\begin{array}{l}\mu\;=\;(r_1+r_2+r_3)/3\\Distance\;=\;(\mu-r_1)^{2}+(\mu-r_2)^{2}+(\mu-r_3)^{2}\\Threshold\;=\;8\\\boldsymbol i\boldsymbol f\;Distance\;\geq\;Threshold:\\\;\;\;\;return\;Law\;expert\;rating\\\boldsymbol e\boldsymbol l\boldsymbol s\boldsymbol e\boldsymbol:\\\;\;\;\;Score\;=\;(\;r_1+r_2+r_3)/3\end{array}\end{array}} $\\
\tblbottomrule 
\end{tabulary}\par 
\end{table*}
Out of 420 pairs of the patents, 85 pairs turn out to be the cases for Law expert rating. After obtaining the 85 pairs rated by the law expert, we combine them with the remaining data, totaling 420 pairs. The mean score of the data is 1.844 with the standard deviation of 1.882.

For the analysis of the technical similarity, we utilize IPC codes in the patent documents. As described in 3.2., we adopt the three-depth structure, meaning that we use the code up to sub-class. For instance, if a patent consists of G06F40/30, G06F40/40, and G06F40/56 codes, we regard the patent as having one IPC code, rather than three. Accordingly, we preprocess the codes. The total number of patents is 840, ranging from 1 to 6. The mean number of IPC codes per patent is 1.602 with a standard deviation of 0.812, as illustrated in Figure~\ref{f-85d554bde98c}.  
\bgroup
\fixFloatSize{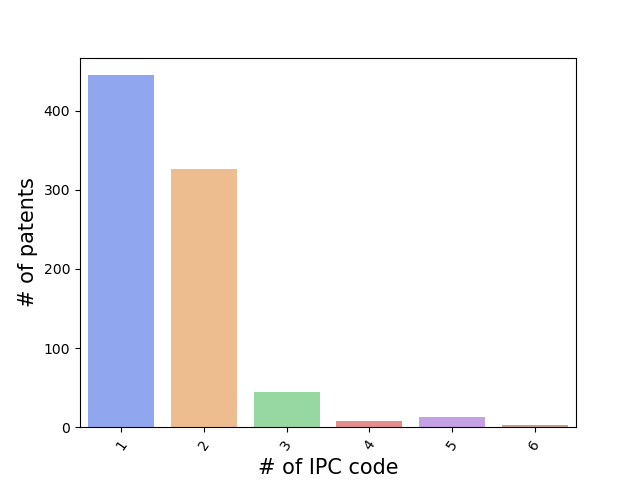}
\begin{figure*}[!htbp]
\centering \makeatletter\IfFileExists{imgs/whole_ipc_code_frequency.png}{\includegraphics[width=.50\linewidth]{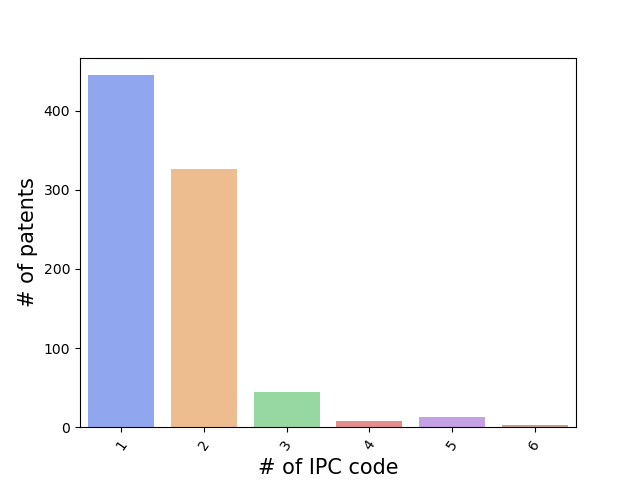}}{}
\makeatother 
\caption{{Frequency of patents depending on the number of IPC codes per patent}}
\label{f-85d554bde98c}
\end{figure*}
\egroup

\subsection{Evaluation metrics}We use Pearson's correlation coefficient and Spearman's one as indicators to evaluate the experimental results. Pearson's correlation coefficient is a measure of the linear correlation between two variables. Pearson's correlation coefficient is calculated by Eq~(\ref{disp-formula-group-95d814cb65444d07bdfd419a3ab9e897}). 
\let\saveeqnno\theequation
\let\savefrac\frac
\def\dispfrac{\displaystyle\savefrac}
\begin{eqnarray}
\let\frac\dispfrac
\gdef\theequation{7}
\let\theHequation\theequation
\label{disp-formula-group-95d814cb65444d07bdfd419a3ab9e897}
\begin{array}{@{}l}\style{font-size:12px}{r=\frac{n\left(\sum_{}^{}xy\right)-\left(\sum_{}^{}x\right)\left(\sum_{}^{}y\right)}{\sqrt{\lbrack n\sum_{}^{}x^{2}-(\sum_{}^{}x)^{2}\rbrack\ast\lbrack n\sum_{}^{}y^{2}-(\sum_{}^{}y){}^{2}\rbrack\;}}}\end{array}
\end{eqnarray}
\global\let\theequation\saveeqnno
\addtocounter{equation}{-1}\ignorespaces 
On the other hand, the Spearman's correlation coefficient is a method of calculating the correlation coefficient using the rank of two values \noindent \noindent rather than the actual value to calculate the correlation coefficient. Spearman's correlation coefficient is obtained byEquation~(\ref{disp-formula-group-13a49fd0ac9f4ffabab334abc3a02dab}).
\let\saveeqnno\theequation
\let\savefrac\frac
\def\dispfrac{\displaystyle\savefrac}
\begin{eqnarray}
\let\frac\dispfrac
\gdef\theequation{8}
\let\theHequation\theequation
\label{disp-formula-group-13a49fd0ac9f4ffabab334abc3a02dab}
\begin{array}{@{}l}\style{font-size:12px}{r=1-\frac{6\sum_{}^{}D^{2}}{n(n-1)}}\end{array}
\end{eqnarray}
\global\let\theequation\saveeqnno
\addtocounter{equation}{-1}\ignorespaces 

\section{Result}In this section, We will explain the process of finding the optimal value when hybridizing the three methods. Afterwards, the experimental results will be explained.

\subsection{Hybrid method Optimizing}
To obtain the weight value in Eq (6), we calculated technical similarity, similarity using bibliographic information, and semantic similarity and then found the optimized value. First, in terms of technical similarity, we find the value of $n$ that maximizes Eq (9) below.

\let\savefrac\frac
\def\dispfrac{\displaystyle\savefrac}
\begin{eqnarray}
\let\frac\dispfrac
\gdef\theequation{9}
\let\theHequation\theequation
\begin{array}{@{}l}\style{font-size:12px}\mathit{Proposed\ Method} = \mathit{Semantic\ Similarity}  \times \left(\frac{\mathit{Technological\ Similarity}}{n}+1\right) 
\end{array}
\end{eqnarray}
\global\let\theequation\saveeqnno
\addtocounter{equation}{-1}\ignorespaces 

We found that it is optimized when the value of n among numbers from 1 to 100 is 9.
\bgroup
\fixFloatSize{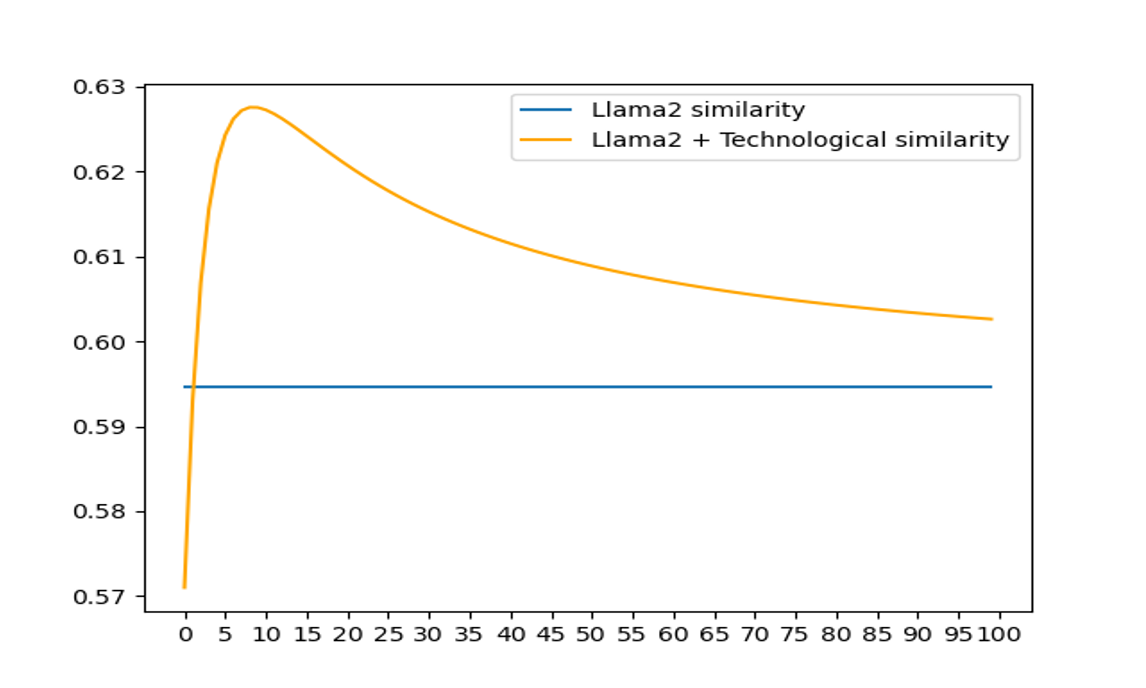}
\begin{figure*}[!htbp]
\centering \makeatletter\IfFileExists{imgs/Opt1.png}{\includegraphics[width=.99\linewidth]{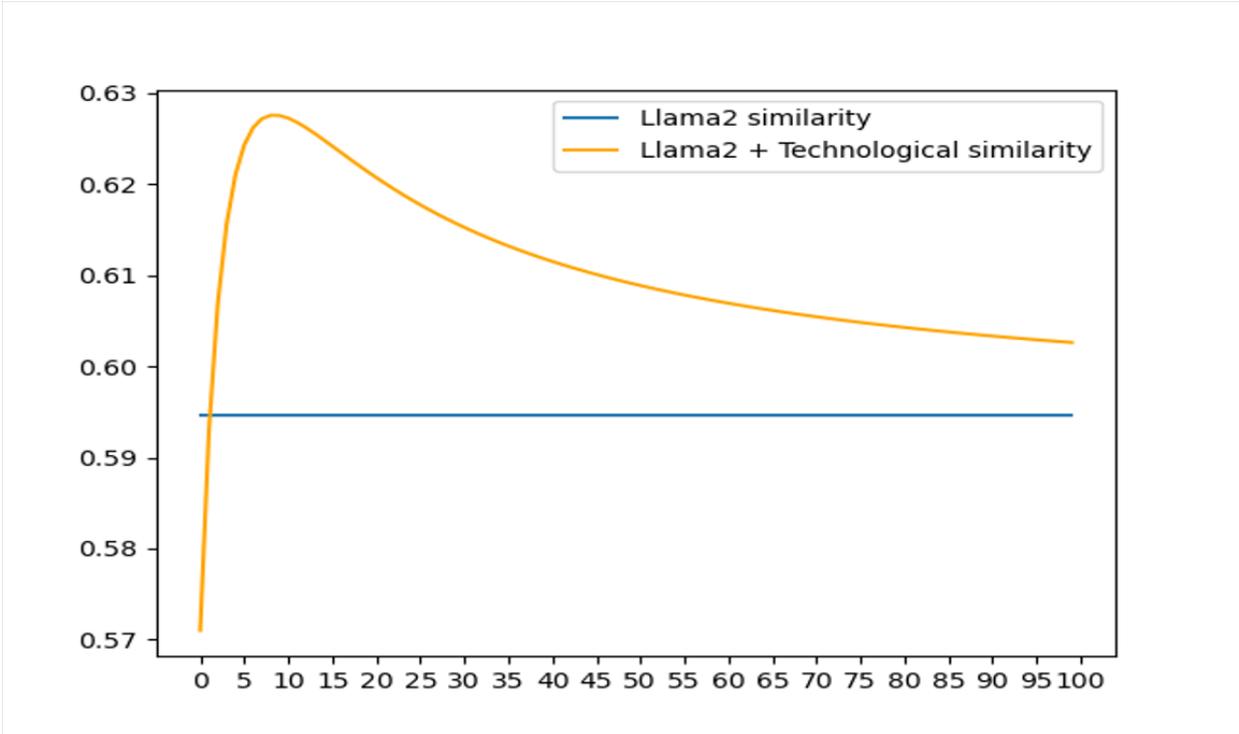}}{}
\makeatother 
\caption{{Optimizing process of semantic similarity and technological similarity}}
\label{f-85d554bde98c}
\end{figure*}
\egroup
Second, in terms of similarity using bibliographic information, we find the value of $m$ that maximizes Eq (10) below.

\let\savefrac\frac
\def\dispfrac{\displaystyle\savefrac}
\begin{eqnarray}
\let\frac\dispfrac
\gdef\theequation{10}
\let\theHequation\theequation
\begin{array}{@{}l}\style{font-size:12px}\mathit{Proposed\ Method} = \mathit{Semantic\ Similarity} \times \left(\frac{\mathit{Bibliography\ Similarity}}{m}+1\right)
\end{array}
\end{eqnarray}
\global\let\theequation\saveeqnno
\addtocounter{equation}{-1}\ignorespaces

We found that it is optimized when the value of m among numbers from 1 to 100 is 8.
\bgroup
\fixFloatSize{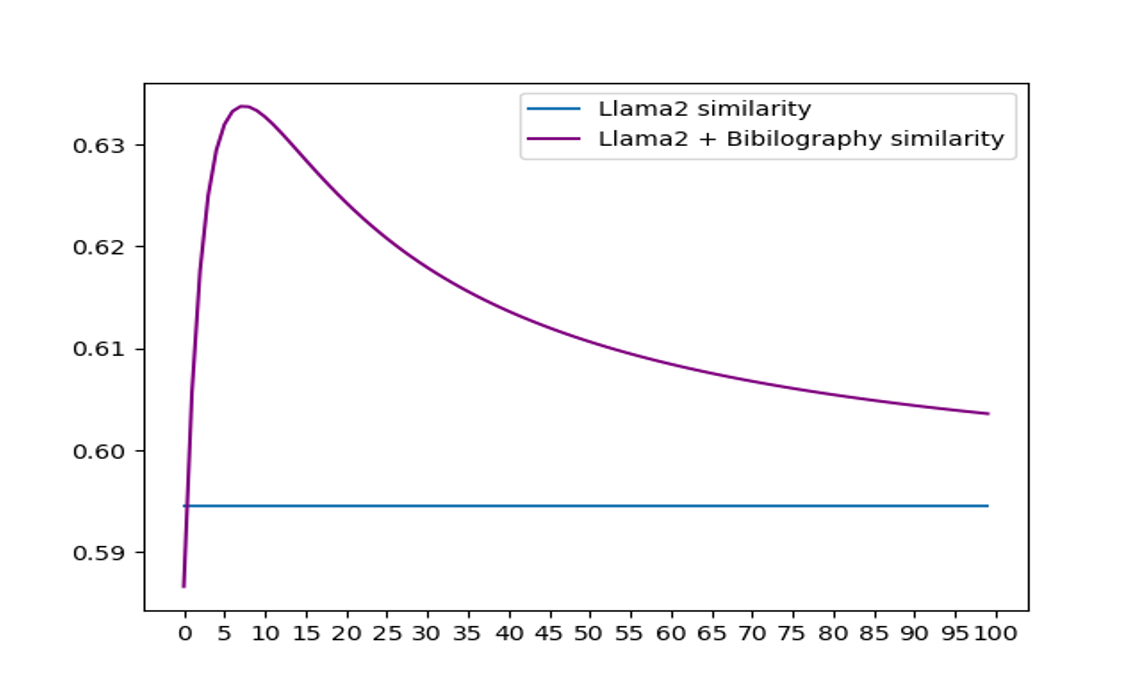}
\begin{figure*}[!htbp]
\centering \makeatletter\IfFileExists{imgs/Opt2.png}{\includegraphics[width=.99\linewidth]{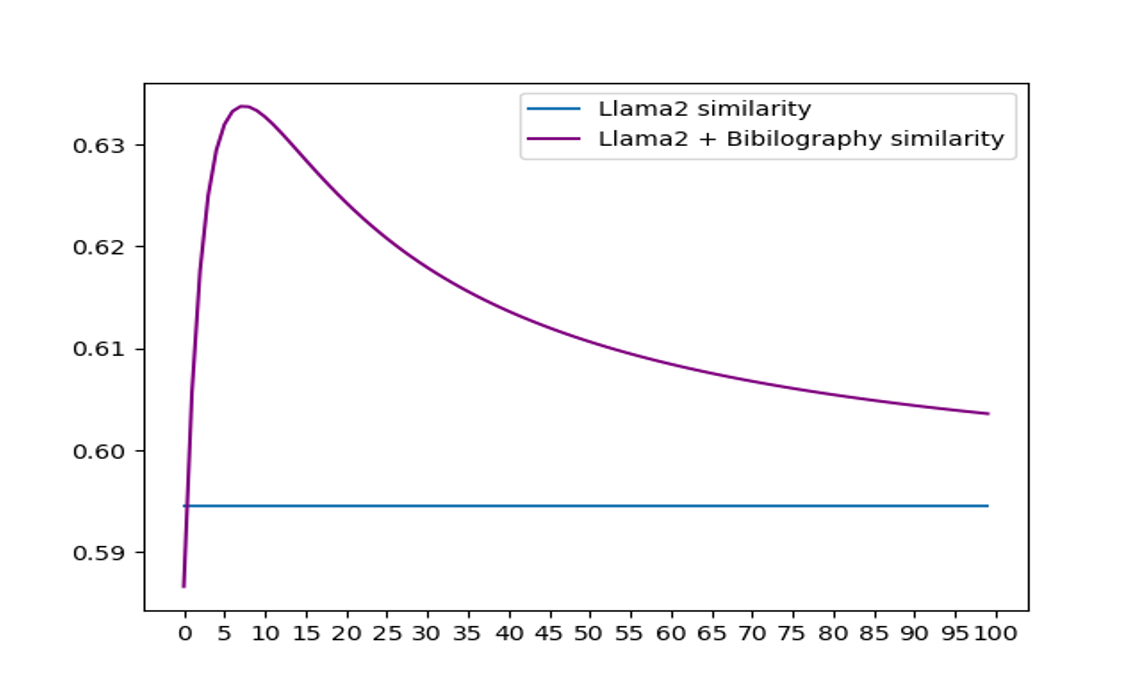}}{}
\makeatother 
\caption{{Optimizing process of semantic similarity and bibliography similarity}}
\label{f-85d554bde98c}
\end{figure*}
\egroup

Therefore, the final Eq (11) of the equation we propose is as follows.

\let\savefrac\frac
\def\dispfrac{\displaystyle\savefrac}
\begin{eqnarray}
\let\frac\dispfrac
\gdef\theequation{11}
\let\theHequation\theequation
\begin{array}{@{}l}\style{font-size:12px}\mathit{Proposed\ Method} = \mathit{Semantic\ Similarity}  \times \left(\frac{\mathit{Technological\ Similarity}}{9}+1\right) \\ \times \left(\frac{\mathit{Bibliography\ Similarity}}{8}+1\right)
\end{array}
\end{eqnarray}
\global\let\theequation\saveeqnno
\addtocounter{equation}{-1}\ignorespaces

\subsection{Result}
We compared the performance of our proposed model with natural language processing models. We evaluated the performance of three traditional deep learning models, BERT, a transformer-based model, a model that considers BERT and technical aspects together, and Llama2 model, a representative LLM model, and our model through Pearson coefficient and Spearman coefficient.
\begin{table*}[!htbp]
\caption{{Model performance}}
\label{tw-67b22344da75}
\def\arraystretch{1}
\ignorespaces 
\centering 
\begin{tabulary}{\linewidth}{p{\dimexpr.4178\linewidth-2\tabcolsep}p{\dimexpr.27139999999999997\linewidth-2\tabcolsep}p{\dimexpr.3108\linewidth-2\tabcolsep}}
\tbltoprule 
\cellcolor[HTML]{E5E5E5}{\textbf{Model}} &
  \cellcolor[HTML]{E5E5E5}{\textbf{Pearson}} &
  \cellcolor[HTML]{E5E5E5}{\textbf{Spearman}}\\
CNN &
  0.1189 &
  0.1362\\
LSTM &
  0.0655 &
  0.0642\\
Bi-LSTM &
  0.1648 &
  0.1669\\
BERT &
  0.4612 &
  0.5246\\
LlaMA2 &
  0.5946 &
  0.6450\\
\textbf{Proposed model} &
  \textbf{0.6342} &
  \textbf{0.6453}\\
\tblbottomrule 
\end{tabulary}\par 
\end{table*}

\section{Discussion}
We propose a hybrid method to measure bibliographic and technical information similarity as well as semantic similarity. The experimental results show that the proposed model has the best performance among the seven models in Table 3. This emphasizes the need to integrate not only semantic similarity but also bibliographic information and technical information similarity when measuring patent similarity. The proposed method can be an effective starting tool for inventions at an early stage by solving a very large problem in patent analysis. 
Many applied studies can be performed using the proposed method. For example, you can respond to the patent troll that has been making headlines recently. Additionally, plagiarism between patents can be analyzed semantically and technically through the proposed similarity measure. It is also possible to identify technology trends by deriving a patent roadmap based on similar patents. Although ours is the best performing model that currently exists, its accuracy is still very low. Therefore, we need to discuss ways to further improve performance. When measuring technical similarity, you should also consider the question of how deep the IPC code should be considered. Also, it is necessary to consider the depth of citation when using bibliographic information. If there is a citation of more than 3 depth, there is a high probability that the same technology will be used, so it is necessary to think about giving additional points.
    
\section{Conclusion}
As one of the basic components of patent analysis, patent similarity measurement can not only efficiently derive technical information, but also detect patent infringement risks and evaluate whether inventions meet novelty and innovation criteria. In a rapidly changing world, methods for measuring similarity between patents still rely on manual classification by experts, with only very important data that must be performed accurately and quickly. A large amount of natural language data measures the similarity between two documents using only natural language processing techniques. However, since patents are documents representing technology made up of legal terms, it is difficult to measure similarity using only natural language processing techniques. Therefore, we proposed a method to measure the similarity between patents using the bibliographic information and technical information of patents as well as natural language processing techniques using the special data characteristics of patents. The effectiveness of the proposed method was verified through experimental results. Our study has scientific significance in that the proposed model showed superior performance to methods using only natural language processing techniques.
    
\section*{Declaration of competing interest}
\noindent The authors declare no conflict of interest.
    
\section*{Funding}
\noindent This work was supported in part by the Business Agency, South Korea under Grant CY220049. 
    
\section*{Acknowledgements}
\noindent The authors acknowledge the helpful comments and suggestions by the anonymous reviewers and the guest editor.
\\
\noindent Author Contributions are as follows:
\noindent Conceptualization, Y.Y.; methodology, Y.Y; validation, S.G., C.J., Y.Y.; formal analysis, S.G., Y.Y; dataset, C.J., Z.S.; Experiment, Y.Y, S.G; writing{\textemdash}original draft preparation, Y.Y., J.L; writing{\textemdash}review and editing, C.J., Y.Y; visualization, S.G., Y.Y; supervision, Y.Y, D.S; All authors have read and agreed to the published version of the manuscript.

\bibliographystyle{elsarticle-num}

\bibliography{refs.bib}

\section*{Author biography}\noindent

\bioItem[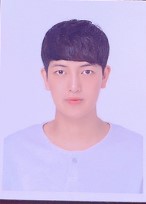]{Yongmin Yoo}{ was born in 1991 in Busan, Republic of Korea. He received a master's degree in Industrial Engineering from Inha university, Korea. He is currently a PhD candidate in University of Auckland, New Zealand. He has experience working as a Natural Language Processing researcher at NHN, one of the biggest companies in Korea. His research interests are in Technological Management, Data Mining, especially Text Mining, Deep Learning, Machine Learning and Natural Language Processing. 

}

\smallskip\noindent 

\bioItem[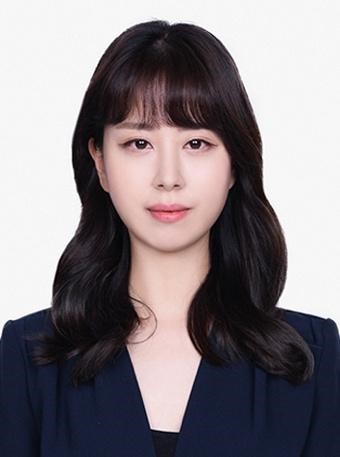]{Cheonkam Jeong}{ was born in 1990 in Seoul, Republic of Korea. She earned both B.A. and M.A. degrees in English Linguistics from Hankuk (Korea) University of Foreign Studies in Seoul. She is currently a PhD candidate in Linguistics at the University of Arizona in the United States. She also holds a M.S. degree in Human Language Technology, which was earned en route to her PhD. She has experience working as a Natural Language Processing Engineer and Speech Specialist for several projects, one of which is sponsored by the U.S. Defense Advanced Research Projects Agency. Her research interests lie in Natural Language Processing and Automatic Speech Recognition/Synthesis.}

\smallskip\noindent 

\bioItem[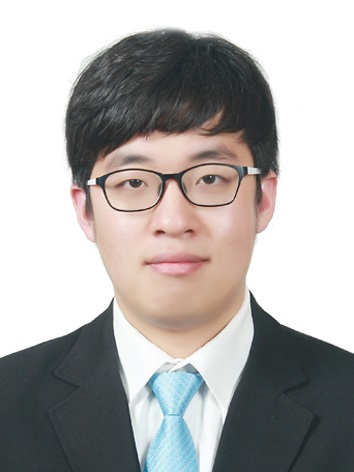]{Sanguk Gim}{ was born in 1991 in Changwon, Republic of Korea. He received a bachelor's degree in Psychology from Kyungnam University. He worked as a Data Analyst at NHN Diquest. He currently works at SR Universe as a Deep Learning Engineer. 

 His research interests are in deep learning, machine learning and natural language processing.

}

\smallskip\noindent 

\bioItem[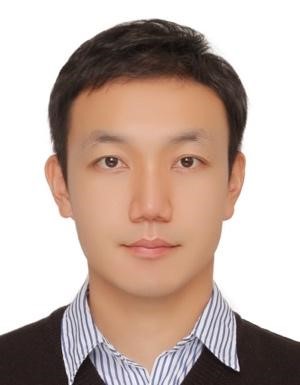]{Junwon Lee}{ was born in 1988 in Gangwon, Republic of Korea. He received a bachelor's degree in Business Administration from Kookmin University. He is currently studying for a master's degree in Data Science at Royal Melbourne Institute of Technology, Melbourne, Australia. He has experience working as a Field Incident Statistical Data Analyst at Group Renault, one of the biggest vehicle companies. His research interests include Data Mining, more specifically Text mining.}

\smallskip\noindent 

\bioItem[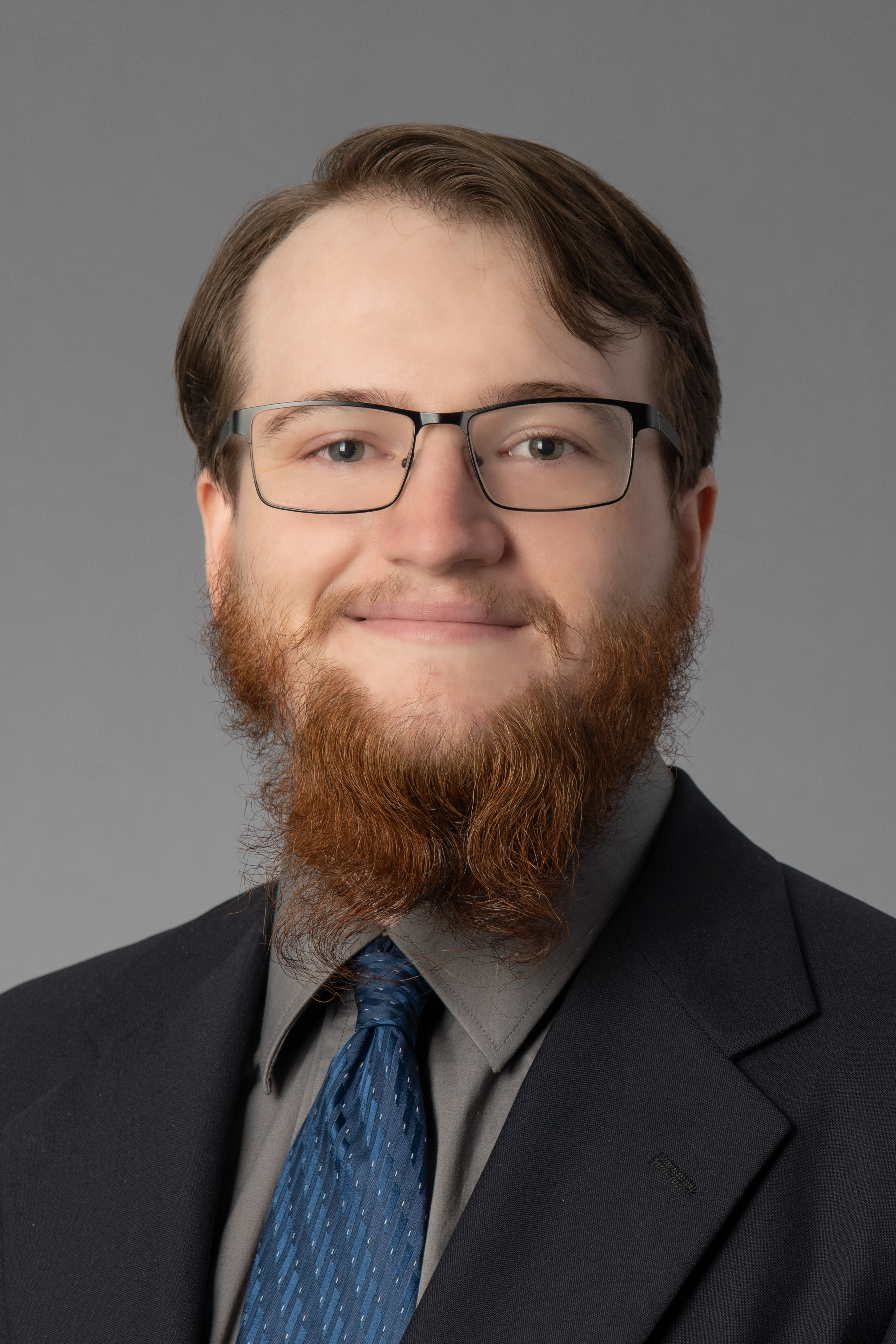]{Zachary Schimke}{ was born in California in the United States.  He earned his B.S. in Mechanical Engineering from California State University Northridge. He is currently a J.D. candidate at the University of Arizona James E. Rogers College of Law. He has been a member of the School's Mock Trial and Patent Moot court teams. He is interested in Intellectual Property as well as Patent Prosecution and Litigation.}

\smallskip\noindent 

\bioItem[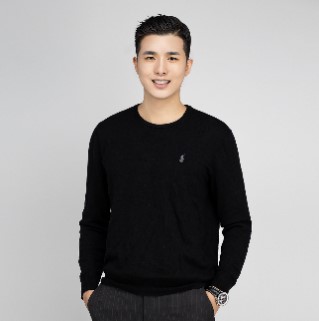]{Deaho Seo}{ was born in 1991 in Seoul, Republic of Korea. He received a B.S degree in Information Systems and a M.S degree in Industrial Engineering from Hanyang University. He is a Ph.D candidate in School of Information, at Yonsei University. He has experience working as a Big Data rSsearcher at the Korea Advanced Institute of Science and Technology and the Korea Electronics Technology Institute. Currently, he is serving as the CEO of Elesther. His research interests include Text Mining, Vision-based Anomaly Detection, Start Factory, and E-commerce solution. He received the Best Young Entrepreneur Award from the Ministry of SMEs and Startups in Korea. He has published 8 books related to artificial intelligence and big data.
}
\printBio 

\end{document}